\documentclass[11pt]{article}
\usepackage{amsfonts,amsmath,amssymb,mathrsfs}
\usepackage[footnotesize]{caption}%
\usepackage{graphicx}

\newcommand{\RRR}{\mathbb{R}}

\begin{document}
\begin{large}
\title{\bf {Quantum Mechanics and imprecise probability}}
\end{large}
\author{Bruno Galvan \footnote{Electronic address: b.galvan@virgilio.it}\\ \small Loc. Melta 40, 38100 Trento, Italy.}
\date{\small November 2007}
\maketitle

\begin{abstract}
An extension of the Born rule, the {\it quantum typicality rule}, has recently been proposed [B. Galvan: Found. Phys. 37, 1540-1562 (2007)]. Roughly speaking, this rule states that if the wave function of a particle is split into non-overlapping wave packets, the particle stays approximately inside the support of one of the wave packets, without jumping to the others.

In this paper a formal definition of this rule is given in terms of {\it imprecise probability}. An imprecise probability space is a measurable space $(\Omega, {\cal A})$ endowed with a {\it set} of probability measures $\cal P$. The quantum formalism and the quantum typicality rule allow us to define a set of probabilities ${\cal P}_\Psi$ on $(X^T, {\cal F})$, where $X$ is the configuration space of a quantum system, $T$ is a time interval and ${\cal F}$ is the $\sigma$-algebra generated by the cylinder sets. Thus, it is proposed that a quantum system can be represented as the {\it imprecise stochastic process} $(X^T, {\cal F}, {\cal P}_\Psi)$, which is a canonical stochastic process in which the single probability measure is replaced by a set of measures. It is argued that this mathematical model, when used to represent macroscopic systems, has sufficient predictive power to explain both the results of the statistical experiments and the quasi-classical structure of the macroscopic evolution.
\end{abstract}

\section{Introduction} \label{int}
Stochastic processes are the standard tools provided by probability theory to represent systems subjected to random evolution. In spite of the fact that quantum mechanics is a probabilistic theory, the presence of quantum interference prevents the standard formalism of quantum mechanics to represent a quantum system as a stochastic process.

In order to understand this, let us consider the most general version of a stochastic process, namely the canonical stochastic process. Let $X$ be the configuration space of a system of particles (for example $\RRR^{3N}$) and $\cal B$ its Borel $\sigma$-algebra. Moreover, let $T$ be a suitable time interval including the origin, and let $X^T$ denote the set of all the trajectories $\lambda: T \rightarrow X$. Given $\Delta \in \cal B$ and $t \in T$, let $(t, \Delta)$ denote the set $\{\lambda \in X^T: \lambda(t) \in \Delta \}$. The sets of this kind will be referred to as {\it s-sets} (an abbreviated name for single-time cylinder-sets); let $\cal S$ denote the class of the s-sets. A {\it canonical stochastic process} is the triple $(X^T, {\cal F}, P)$, where ${\cal F}$ is the $\sigma$-algebra generated by the s-sets (or equivalently, by the cylinder sets) and $P$ is a probability measure on ${\cal F}$. According to the Kolmogorov reconstruction theorem, the probability $P$ is univocally determined by its {\it finite dimensional distributions}, i.e. by its values at the finite intersections of s-sets:
\begin{equation} \label{r1}
P[(t_1, \Delta_1) \cap ... \cap (t_n, \Delta_n)].
\end{equation}
Since $P$ is a probability measure, this expression is additive, i.e., if $\Delta_i \cap \Delta'_i= \emptyset$, we have 
$$
P[ \ldots \cap (t_i, \Delta_i \cup \Delta'_i )\cap \ldots ] = P[ \ldots \cap (t_i, \Delta_i)\cap \ldots ] + P[ \ldots \cap (t_i, \Delta'_i )\cap \ldots ].
$$
The physical interpretation of a canonical stochastic process is simple: the evolution of the system of particles during the time interval $T$ is represented by a trajectory chosen at random from $X^T$.

Let us now attempt to represent a system of quantum particles as a canonical stochastic process. In the quantum case as well, we can take $X$ as the configuration space and, for the time being, let us assume that also in the quantum case the particles follow a trajectory of $X^T$. The problem is then to define the probability $P$, i.e. to find a quantum expression for the finite dimensional distributions (\ref{r1}).

We know that a normalized wave function $\Psi(t)=U(t)\Psi_0$ is associated with the quantum system, where $U(t)$ is the unitary time evolution operator and $\Psi_0$ is the wave function of the system at the time $t=0$. According to the Born rule, the probability of finding the particles in the region $\Delta \in \cal B$ at the time $t$ is $||E(\Delta)\Psi(t)||^2$, where $E(\cdot)$ is the projection valued measure on $\cal B$. In other words, we can say that $||E(\Delta)\Psi(t)||^2$ is the probability that a trajectory chosen at random from $X^T$ belong to the s-set $(t, \Delta)$. Thus for $n=1$ we have a valid quantum expression for (\ref{r1}), namely
\begin{equation}
P[(t, \Delta)]=||E(\Delta)\Psi(t)||^2.
\end{equation}
The problems arise when $n > 1$. A tentative quantum expression for (\ref{r1}) could be
\begin{equation} \label{r2}
||E(\Delta_n)U(t_n-t_{n-1})E(\Delta_{n-1}) \ldots U(t_2-t_1) E(\Delta_1) \Psi(t_1)||^2,
\end{equation}
where the assumption is made that $t_1 \leq \ldots \leq t_n$. According to the Born rule and to the reduction postulate, (\ref{r2}) is the probability of finding the particles in the regions $\Delta_i$ at the times $t_i$, for $i=1, \ldots, n$. However (\ref{r2}) is not an admissible expression for the finite dimensional distributions, because it is not additive, i.e. if $\Delta_i \cap \Delta'_i = \emptyset$, in general we have 
$$
||\ldots E(\Delta_i \cup \Delta'_i) \ldots \Psi(t_1)||^2 \neq ||\ldots E(\Delta_i) \ldots \Psi(t_1)||^2 + ||\ldots E(\Delta'_i) \ldots \Psi(t_1)||^2.
$$
The non additivity of (\ref{r2}) corresponds to the well known interference phenomena of quantum mechanics. Another possible expression, namely:
\begin{equation}
Re \langle \Psi(t_n) |E(\Delta_n)U(t_n-t_{n-1})E(\Delta_{n-1}) \ldots E(\Delta_1) | \Psi(t_1) \rangle;
\end{equation}
is also not admissible, because, although it is additive, it is non-positive definite. We must therefore conclude that a valid expression for the finite dimensional distributions, and therefore a probability measure for $X^T$, cannot be extracted from the standard quantum formalism.

Two possible solutions to this problem can be proposed, corresponding to different formulations/interpretations of quantum mechanics. The first is to simply remove the set $X^T$, and to let the wave function be the only mathematical entity representing a quantum system. This is, for example, the position of the Copenhagen and of the Many Worlds interpretations \cite{everett}. For example, in \cite{heisenberg} Heisenberg explicitly connects the quantum interference phenomena and the necessity to renounce a description of the motion of the particles in terms of trajectories. The second solution is to add a new element to the standard quantum formalism which allows us to define the required probability measure. This is the case, for example, of Nelson's stochastic mechanics \cite{nelson}, which introduces a stochastic differential equation, and of Bohmian mechanics \cite{allori}, which introduces the guidance equation\footnote{In \cite{peruzzi} it is shown that the guidance equation is the limiting case of a general class of stochastic differential equations which also include Nelson's theory.}. For various reasons that we do not discuss here, for many physicists none of these formulations is satisfying.

In this paper a third solution is proposed, which maintains the set $X^T$ and does not add new elements to the quantum formalism, but rather admits the possibility that the quantum formalism defines a {\it set of probability measures} on $X^T$ instead of a single probability. For example, if $\cal M$ is the set of all the probability measures on $(X^T, {\cal F})$, the Born rule defines the set of probabilities
\begin{equation}
{\cal P}_{\Psi B}: =\{P \in {\cal M}: P[(t, \Delta)]=||E(\Delta)\Psi(t)||^2 \; \; \hbox{for all} \; \; (t, \Delta) \in {\cal S}\}.
\end{equation}
This paper therefore proposes that, instead of a ``precise'' stochastic process $(X^T, {\cal F}, P)$, the correct mathematical model of a quantum system is an ``imprecise'' stochastic process $(X^T, {\cal F}, {\cal P}_\Psi)$, where ${\cal P}_\Psi$ is a suitable set of probability measures. The word ``imprecise'' has been used intentionally, because
{\it imprecise probability} is a generic term which also includes the theory of sets of probabilities \cite{imprecise}. The meaning of this representation is that a possible evolution of the system is represented by a trajectory chosen at random from $X^T$ {\it according to any one of the probabilities of $P_\Psi$}. In general, in an imprecise stochastic process there are events without a well defined probability, and therefore the predictive power of the process is limited to those events $A$ for which $P(A)$ has, at least approximately, the same value for all the probabilities $P \in {\cal P}_\Psi$. 

A fundamental element of the proposed solution is the use of the quantum typicality rule instead of the Born rule to define the set ${\cal P}_\Psi$. Let us explain this. The set ${\cal P}_{\Psi B}$ defined by the Born rule arguably explains the results of the statistical experiments, because it attributes a well defined probability to any set of configurations at a given time. However, the Born rule does not establish any correlation between the positions of the particles at different times, and therefore it cannot define a dynamical structure for the trajectories. Therefore, if the imprecise process $(X^T, {\cal F}, {\cal P}_{B\Psi})$ is used to represent a macroscopic system, for example the universe, it cannot explain the quasi-classical structure of the macroscopic evolution. This is also a well known problem in connection with the Copenhagen and the Many Words interpretations.

In a recent paper we proposed a new quantum rule, the {\it quantum typicality rule}, according to which, roughly speaking, the particles follow the branches of the wave function \cite{galvan}. In the present paper a set ${\cal P}_\Psi$ of probabilities corresponding to the quantum typicality rule is defined, and some of its properties are studied. This set is contained in ${\cal P}_{\Psi B}$, i.e. the quantum typicality rule implies the Born rule. Moreover, since the quantum typicality rule establishes a correlation between the positions of the particles at two different times, the set ${\cal P}_\Psi$ arguably explains the macroscopic quasi-classical structure of the trajectories.

\vspace{3mm}
The paper is structured as follows: in section \ref{quantum} the quantum typicality rule is reviewed. In section \ref{theory} a short review of the theory of imprecise probability is given. In section \ref{born} a formal definition of the Born rule in terms of imprecise probability is given. In section \ref{thequantum} a formal definition of the quantum typicality rule in terms of imprecise probability is given, and the {\it quantum process}, i.e. the proposed mathematical model of a quantum system, is defined. In section \ref{two} two properties of quantum processes are studied. Section \ref{discussion} presents a concluding discussion about the formulation of quantum mechanics based on the quantum typicality rule. 

\section{The quantum typicality rule} \label{quantum}
Let us first introduce the notion of typicality in a probability space. Let $(\Omega, {\cal A}, P)$ be a probability space, and let $A$ and $B$ be two measurable subsets of $\Omega$, with $P(B) \neq 0$. The set $A$ is said to be {\it typical} relative to $B$ if 
\begin{equation}
\frac{P(A \cap B)}{P(B)} \approx 1,
\end{equation}
where $\approx 1$ is understood to mean $\geq 1 - \epsilon$, with $\epsilon \ll 1$. If $A$ is typical relative to $B$, then the overwhelming majority of the elements of $B$ also belongs to $A$. From the empirical point of view, the consequence of the typicality of $A$ is that a single element chosen at random from $B$ will also belong to $A$. Two sets $A$ and $B$ are said to be {\it mutually typical} if $A$ is typical relative to $B$ and vice-versa. Mutual typicality can be expressed by the condition
\begin{equation} \label{3}
\frac{P(A \cap B)}{\max\{P(A), P(B)\}} \approx 1.
\end{equation}

The notion of typicality is used in Bohmian mechanics in order to prove the {\it quantum equilibrium hypothesis} \cite{durr1} and it is also (implicitly) the basis for Boltzmann's derivation of the second law of thermodynamics \cite{goldstein}.

\vspace{3mm}
Let us now introduce the quantum typicality rule. In its simplest and most intuitive form, the quantum typicality rule states: suppose that the wave function of a particle is the sum of two non-overlapping wave packets. Then, during the time over which the wave packets are non-overlapping, the particle stays inside the support of one of the two wave packets, without jumping to the other one.

For example, let us consider the following simple experiment:
\begin{center}
\includegraphics {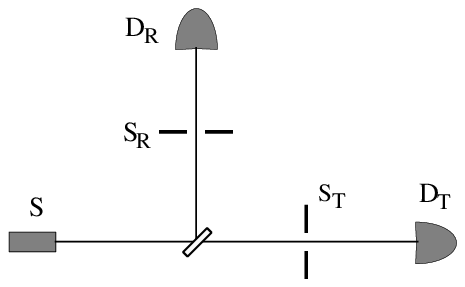} \\
Fig.1
\end{center}

The source $S$ emits photons towards a beam splitter, the reflected (transmitted) photons are detected by the detector $D_T$ ($D_R$), and $S_R$ and $S_T$ are two slits. The quantum typicality rule states that the photons detected, for example, by the detector $D_R$, cross the slit $S_R$. Even if this assumption is very reasonable, it cannot be deduced from standard quantum mechanics, which does not predict the trajectory of a quantum system between the preparation and the measurement times. This feature of quantum mechanics is essentially the origin of its difficulty in explaining the emergence of a quasi-classical world.

We can easily express the rule in a mathematical form. Let $\Psi(t)=U(t)\Psi_0$ be the wave function of a particle (or of a system of particles). Let us suppose that at a time $t_1$ the wave function can be expressed as the sum of two non-overlapping wave packets $\phi$ and $\phi_\perp=\Psi(t_1)-\phi$, and that at a time $t_2 > t_1$ the two wave packets are still non-overlapping, i.e. $U(t_2-t_1)\phi$ and $U(t_2-t_1)\phi_\perp$ are non-overlapping. This implies that two subsets $\Delta_1$ and $\Delta_2$ of the configuration space of the particle exist, such that \begin{equation} \label{cc1}
\phi \approx E(\Delta_1)\Psi(t_1) \; \; \hbox{and} \; \; U(t_2-t_1)\phi \approx E(\Delta_2)\Psi(t_2).
\end{equation}
Due to the unavoidable spreading of the wave function, the wave packets can be only approximately non-overlapping. This is the reason for using the approximate equality symbol in (\ref{cc1}). The sets $\Delta_1$ and $\Delta_2$ can be considered as the supports of $\phi$ and $U(t_2-t_1)\phi$, respectively. The conditions (\ref{cc1}) can be combined to give the condition
\begin{equation} \label{cc2}
U(t_2-t_1)E(\Delta_1)\Psi(t_1) \approx E(\Delta_2)\Psi(t_2).
\end{equation}
This reasoning can also be reversed: given two subsets $\Delta_1$ and $\Delta_2$ satisfying condition (\ref{cc2}), the wave packet $\phi:= E(\Delta_1)\Psi(t_1)$ satisfies the conditions of (\ref{cc1}). 

The quantum typicality rule states that, if the particle is in $\Delta_2$ at the time $t_2$ and condition (\ref{cc2}) holds, then the particle was in $\Delta_1$ at the time $t_1$. Since the two times are symmetric, it is natural to assume also the reverse conclusion: if the particle is in $\Delta_1$ at the time $t_1$ then it will be in $\Delta_2$ at the time $t_2$.

\vspace{3mm}
Let us take a further step and assume, as proposed in the introduction, that a quantum particle follows a trajectory belonging to $X^T$. Note that this assumption is implicitly contained in the intuitive formulation of the quantum typicality rule, because such a rule assumes that the particle has a position at a suitable time even if no measurement is performed at that time. Then the quantum typicality rule can be expressed by stating that condition (\ref{cc2}) implies that the two s-sets $(t_1, \Delta_1)$ and $(t_2, \Delta_2)$ are mutually typical.

A more compact notation can also be introduced. If $S$ denotes the s-set $(t, \Delta)$, let $\Psi(S)$ denote the state $U^\dagger(t)E(\Delta)U(t)\Psi_0$. With this notation, condition (\ref{cc2}) assumes the form 
\begin{equation} \label{cc3}
||\Psi(S_1) - \Psi(S_2)||^2 \approx 0,
\end{equation}
where $S_1=(t_1, \Delta_1)$ and $S_2=(t_2, \Delta_2)$. The norm has been squared for reasons that will become clear in section \ref{thequantum}. Due to the presence of the approximate equality, condition (\ref{cc3}) must be normalized. In order to simplify the normalization, we impose the further natural condition that $||\Psi(S_1)||=||\Psi(S_2)||$. In conclusion, the quantum typicality can be expressed as follows:
\begin{trivlist}
\item[\hspace\labelsep{\bf Quantum Typicality Rule:}] if $S_1$ and $S_2$ are two s-sets such that $||\Psi(S_1)||=||\Psi(S_2)||$ and 
\begin{equation} \label{4}
\frac{||\Psi(S_1) - \Psi(S_2)||^2}{||\Psi(S_1)||^2} \ll 1,
\end{equation}
than $S_1$ and $S_2$ are mutually typical.
\end{trivlist}
The constraint $||\Psi(S_1)||=||\Psi(S_2)||$, which was not present in the first formulation of the rule \cite{galvan}, will be discussed in section \ref{thequantum}.

The problem is now to correlate the definition of mutual typicality given by the probabilistic expression (\ref{3}) with the one given by the quantum expression (\ref{4}). This correlation is conceptually similar to the correlation $P(S)=||E(\Delta)\Psi(t)||^2$ given by the Born rule, and it will be realized by means of the theory of imprecise probability.

\section{Theory of imprecise probability} \label{theory}

A very short review is given here of the theory of sets of probabilities, which is a part of the theory of imprecise probability \cite{huber}. 

Let $(\Omega, {\cal A})$ be a measurable space and $\cal M$ be the set of all the probability measures on $(\Omega, {\cal A})$. Let $\cal P$ be an arbitrary non-empty subset of $\cal M$. The {\it upper} and the {\it lower probability} induced by $\cal P$ are the two set functions $P_*, P^*:{\cal A} \rightarrow \RRR_+$ defined by:
\begin{equation}
P_*(A)=\inf_{P \in \cal P} \, P(A); \; \; \; P^*(A)=\sup_{P \in \cal P} \, P(A).
\end{equation}
We can easily see that $P_*$ and $P^*$ satisfy the following properties:
\begin{eqnarray}
& & 0 \leq P_*(A) \leq P^*(A) \leq 1: \label{p1} \\
&& P_*(\emptyset)=P^*(\emptyset)=0; \; \; P_*(\Omega)=P^*(\Omega)=1; \label{p2} \\
& & P_*(A) + P^*(A^c) = 1; \label{p3} \\
& & P_*(A \cup B)  \geq P_*(A) + P_*(B) \; \; \hbox{for} \; \; A \cap B = \emptyset;  \label{p4} \\
& & P^*(A \cup B)  \leq P^*(A) + P^*(B) \; \; \hbox{for} \; \; A \cap B = \emptyset; \label{p5}\\
& & P_*(A) \leq P_*(B)\; \; \hbox{and} \; \; P^*(A) \leq P^*(B) \; \; \hbox{for} \; \; A \subseteq B. \label{p6}
\end{eqnarray}
Equation (\ref{p3}) states that $P_*$ and $P^*$ are conjugate; equations (\ref{p4}) and (\ref{p5}) state that $P_*$ is superadditive and $P^*$ is subadditive; equation (\ref{p6}) states that $P_*$ and $P^*$ are monotone.

The triple $(\Omega, {\cal A}, {\cal P})$ will be referred to as an {\it imprecise probability space}. In the case in which $\Omega=X^T$ and ${\cal A}={\cal F}$, the more specific term {\it imprecise stochastic process}  will be used. The predictive power of an imprecise probability space is limited to those events $A$ for which $P(A)$ assumes approximately the same value for all $P \in {\cal P}$. Such a condition is satisfied for example if a positive number $a$ exists such that 
\begin{equation}
\frac{|P(A)- a|}{a} \ll 1 \; \; \hbox{for all} \; \; P \in {\cal P}.
\end{equation}

\vspace{3mm}
Let us now study how sets of probability measures can be defined. Let $\cal D$ be an arbitrary subset of $\cal F$, and $f_*: {\cal D} \rightarrow R_+$ a non negative set function. Let us define a set of probability measures $\cal P$ as
\begin{equation} \label{t}
{\cal P}:=\{ P \in {\cal M}: P(A) \geq f_*(A) \; \; \hbox{for all} \; \; A \in {\cal D}\}.
\end{equation}
Alternatively, the set $\cal P$ can be defined as
\begin{equation} \label{tt}
{\cal P}:=\{ P \in {\cal M}: P(B) \leq f^*(B) \; \; \hbox{for all} \; \; B \in {\cal D}^c\},
\end{equation}
where ${\cal D}^c:= \{ B : B^c \in {\cal D}\}$, and $f^*(B):=1 - f_*(B^c)$.

\vspace{3mm}
We now have the following lemma \cite{huber}: the set $\cal P$ defined by (\ref{t}) is not empty iff, for any pair of finite sequences $\{a_1, \ldots,  a_n\}$ and $\{A_1, \ldots, A_n\}$ of non negative numbers and of sets of $\cal D$, the condition
\begin{equation} \label{t2}
\sum_{i=1}^n a_i {\bf 1}_{A_i} (\omega) \leq 1 \; \; \forall \omega \in \Omega
\end{equation}
implies the condition
\begin{equation}
\sum a_i f_*(A_i) \leq 1,
\end{equation}
where ${\bf 1}_{A_i}$ is the characteristic function of the set $A_i$ (see note\footnote{Actually, the proof in the given reference applies only to the case of a finite set $\Omega$.}).

\section{The Born process} \label{born}
Let us first apply the theory of imprecise probability to the Born rule. As mentioned in the introduction, the Born rule defines the set of probabilities on $(X^T, {\cal F})$ satisfying the condition $P(S)=||\Psi(S)||^2$ for any s-set $S$.

By referring to the notation used in the previous section, we have the condition that $(\Omega, {\cal A}) = (X^T, {\cal F})$, ${\cal D} = \cal S$ and $f_*(S) = ||\Psi(S)||^2$. Thus the set ${\cal P}_{\Psi B}$ defined by the Born rule is
\begin{equation}
{\cal P}_{\Psi B}:=\{ P \in {\cal M}: P(S) \geq ||\Psi(S)||^2 \; \; \hbox{for all} \; \; S \in {\cal S}\}.
\end{equation}
Note that ${\cal S}^c={\cal S}$ and $f^*=f_*$. We therefore have $P_*(S)=P^*(S)$ for all $S \in \cal S$, which implies that $P(S)=||\Psi(S)||^2$ for all $S \in \cal S$ and for all $P \in {\cal P}_{\Psi B}$, as required. The class ${\cal P}_{\Psi B}$ is not empty because it contains at least the probability $P$ defined by the finite dimensional distributions
$$
P(S_1 \cap \ldots \cap S_n):=||\Psi(S_1)||^2 \ldots ||\Psi(S_n)||^2,
$$
where the assumption is made that $t_i \neq t_j$ for $i \neq j$.

The imprecise process $(X^T, {\cal F}, {\cal P}_{\Psi B})$ will be referred to as the {\it Born process}.

\section{The quantum process} \label{thequantum}

Let us now attempt to define a set of probabilities ${\cal P}_\Psi$ corresponding to the quantum typicality rule. The most natural definition appears to be the following: the set $\cal D$ is 
\begin{equation}
{\cal D}:= \{ S_1 \cap S_2 : S_1, S_2 \in {\cal S}, \; S_1\cap S_2 \not \in {\cal S}, \; ||\Psi(S_1)||=||\Psi(S_2)|| \},
\end{equation}
and the set function $f_*$ is
\begin{equation}
f_*(S_1 \cap S_2) =||\Psi(S_1)||^2 -  ||\Psi(S_1) - \Psi(S_2)||^2,
\end{equation}
The s-sets have been excluded from $\cal D$ because $f_*$ is not a well defined set function for $S_1 \cap S_2 \in \cal S$. The set ${\cal P}_\Psi$ is then defined as
\begin{equation}
{\cal P}_\Psi:=\{ P \in {\cal M}: P(S_1 \cap S_2) \geq ||\Psi(S_1)||^2 - ||\Psi(S_1) - \Psi(S_2)||^2 \; \; \hbox{for all} \; \; S_1 \cap S_2 \in {\cal D} \}. \end{equation}

Let us introduce the conjugate elements $f^*$ and ${\cal D}^c$:
\begin{equation}
{\cal D}^c:= \{ S_1 \cup S_2 : S_1, S_2 \in {\cal S}, \; S_1\cup S_2 \not \in {\cal S},  \; ||\Psi(S_1)||=||\Psi(S_2)|| \};
\end{equation}
\begin{equation}
f^*(S_1 \cup S_2) = ||\Psi(S_1)||^2 + ||\Psi(S_1) - \Psi(S_2)||^2.
\end{equation}
We can easily see that
\begin{eqnarray}
\sup_{\{S_2: \, S_1 \cap S_2 \in \cal D\}} & & f_*(S_1 \cap S_2) = ||\Psi(S_1)||^2, \label{c1} \\ 
\inf_{\{S_2: \, S_1 \cup S_2 \in {\cal D}^c \}} & & f^*(S_1 \cup S_2) = ||\Psi(S_1)||^2. \label{c2}
\end{eqnarray}
Since
\begin{equation}
f_*(S_1 \cap S_2) \leq P(S_1 \cap S_2 ) \leq P(S_1) \leq P(S_1 \cup S_2) 
\leq f^*(S_1 \cup S_2),
\end{equation}
we obtain the condition $P(S)=||\Psi(S)||^2$ for all $S \in \cal S$ and for all $P \in {\cal P}_\Psi$. Thus ${\cal P}_\Psi \subset  {\cal P}_{\Psi B}$, i.e. the formal quantum typicality rule implies the Born rule. For this reason the condition 
\begin{equation} \label{ww}
P(S_1 \cap S_2) \geq ||\Psi(S_1)||^2 - ||\Psi(S_1)-\Psi(S_2)||^2 \; \; \hbox{for} \; \; ||\Psi(S_1)||^2=||\Psi(S_2)||^2
\end{equation}
is also satisfied for $S_1 \cap S_2 \in \cal S$. Indeed in this case we have 
$$
P(S_1 \cap S_2)=||\Psi(S_1 \cap S_2)||^2 \geq 2 ||\Psi(S_1 \cap S_2)||^2 - ||\Psi(S_2)||^2 = ||\Psi(S_1)||^2 - ||\Psi(S_1) - \Psi(S_2)||^2.
$$

The imprecise process $(X^T, {\cal F}, {\cal P}_\Psi)$ will be referred to as a {\it quantum process}, and the defining condition (\ref{ww}) will be referred to as the {\it formal} quantum typicality rule. The adjective ``formal'' has been adjoined in order to distinguish condition (\ref{ww}) from the quantum typicality rule as expressed in section \ref{quantum}, which will be referred to as the {\it physical} quantum typicality rule. These adjectives will, however, be omitted when not required for reasons of clarity.

The physical quantum typicality rule can be derived trivially from the formal rule. Thus, if $S_1$ and $S_2$ be two s-sets such that $||\Psi(S_1)||=||\Psi(S_2)||$ and 
$$
\frac{||\Psi(S_1) - \Psi(S_2)||^2}{||\Psi(S_1)||^2} \leq \epsilon \ll 1,
$$
than 
$$
\frac{P(S_1 \cap S_2)}{\max\{P(S_1), P(S_2)\}}=\frac{P(S_1 \cap S_2)}{||\Psi(S_1)||^2} \geq 
\frac{||\Psi(S_1)||^2 - ||\Psi(S_1) - \Psi(S_2)||^2}{||\Psi(S_1)||^2} \geq 1 - \epsilon
$$
for all $P \in {\cal P}_\Psi$.

\vspace{3mm}
The definition of the formal quantum typicality rule given by (\ref{ww}) appears to be the simplest and most natural one, corresponding to the physical quantum typicality rule. However, some variants of this definition are possible, which we will now examine.

Let us first discuss the constraint
\begin{equation} \label{co}
||\Psi(S_1)||^2=||\Psi(S_2)||^2.
\end{equation}
Such a constraint can probably be removed from both the formal and the physical formulations of the rule. In this case condition (\ref{ww}) must be replaced by the condition
\begin{equation}
P(S_1 \cap S_2) \geq \min\{||\Psi(S_1)||^2,||\Psi(S_2)||^2\} - ||\Psi(S_1)-\Psi(S_2)||^2.
\end{equation}
If  ${\cal P}'_{\Psi}$ is the corresponding set of probabilities, we have ${\cal P}'_{\Psi} \subseteq {\cal P}_\Psi$. Thus, the constraint (\ref{co}) actually gives rise to a more general set of probabilities. This fact, together with the fact that this simplifies both the formal and the physical formulations of the quantum typicality rule, suggests that the constraint (\ref{co}) is appropriate.

Here it should be noted that the formal and the physical formulations are not totally equivalent, because the latter acts only in the typicality regime, i.e. when 
\begin{equation} \label{ry}
\frac{||\Psi(S_1) - \Psi(S_2)||^2}{||\Psi(S_1)||^2} \ll 1.
\end{equation}

On the contrary, condition (\ref{ww}) also imposes a constraint on the probability when (\ref{ry}) does not hold true. In order to eliminate such a difference, condition (\ref{ww}) can be replaced by
\begin{eqnarray} 
& & P(S_1 \cap S_2) \geq ||\Psi(S_1)||^2 - ||\Psi(S_1)-\Psi(S_2)||^2 \; \; \hbox{for} \; \;  ||\Psi(S_1)||=||\Psi(S_2)|| \; \; \hbox{and} \label{sup} \\
& &  ||\Psi(S_1)-\Psi(S_2)||^2  \leq \epsilon ||\Psi(S_1)||^2, \nonumber
\end{eqnarray}
where $\epsilon$ is a suitable ``small'' positive number. If ${\cal P}''_{\Psi}$ is the set of probabilities defined by this condition, then  ${\cal P}_\Psi \subseteq {\cal P}''_{\Psi}$. The problem with this definition is its vagueness, because a precise value for $\epsilon$ cannot be provided. Note also that also a condition of the type 
\begin{equation}
P(S_1 \cap S_2) \geq ||\Psi(S_1)||^2 - \alpha ||\Psi(S_1) - \Psi(S_2)||^2 \; \; \hbox{for} \; \;  ||\Psi(S_1)||=||\Psi(S_2)||,
\end{equation}
where $\alpha$ is positive number not ``too small'' and not ``too big'', could be consistent with the physical quantum typicality rule. Also this definition is vague. 

There are conceptual reasons which suggest that the definition of the set ${\cal P}_\Psi$ is necessarily vague, in the sense that slightly different definitions of ${\cal P}_\Psi$ are empirically indistinguishable. These reasons are connected with the fact that we have access to the past structure of the trajectories only through the memories of the past which are encoded in the present configuration of our recording devices. See \cite{galvan} for a discussion of this point. This subject will be further developed in a future paper.

\vspace{3mm}
Unfortunately we cannot make any statement about the consistency of the quantum typicality rule, i.e. we cannot prove that the set ${\cal P}_\Psi$ is not empty. The problem of the consistency of the quantum typicality rule was also discussed in \cite{galvan}, where some inequalities making such a consistency plausible were proposed. Here the problem has not yet been solved in a rigorous way, but at least it has been formulated in a precise way.

\section{Two properties of quantum processes} \label{two}

Let us study two properties of quantum processes. 

In general, the probability of the intersection of two non equal time s-sets $S_1$ and $S_2$ is not well defined by ${\cal P}_\Psi$, i.e. $P(S_1 \cap S_2)$ may have different values for different $P \in {\cal P}_\Psi$. For example, let the wave function $\Psi(t)$ be a single wave packet, with $\Delta_1$ and $\Delta_2$ such that $||E(\Delta_1)\Psi(t_1)||^2=||E(\Delta_2)\Psi(t_2)||^2=1/2$. If $|t_2-t_1|$ is large enough, we have in any case (that is also in the case in which $\Delta_1 \approx \Delta_2$) that $||E(\Delta_2)\Psi(t_2)- U(t_2 - t_1)E(\Delta_1)\Psi(t_1)||^2 \not \approx 0$, and therefore there is no constraint preventing $P[(t_1, \Delta_1) \cap (t_2, \Delta_2)]$ from assuming a wide range of values.

There is however a typical situation in which $P(S_1 \cap S_2)$ has (approximately) the same value for all $P \in {\cal P}_\Psi$. Let $\phi(t):=U(t-t_1)\phi$ be a wave packet which does not overlap $\phi_\perp(t):=\Psi(t) - \phi(t)$ at the times $t_1$ and $t_2$ and let $\Delta_1$ and $\Delta_2$ be the supports of $\phi(t_1)$ and $\phi(t_2)$, respectively, with $||E(\Delta_1)\Psi(t_1)||=||E(\Delta_2)\Psi(t_2)||$. According to the quantum typicality rule, $S_1=(t_1, \Delta_1)$ and $S_2=(t_2, \Delta_2)$ are mutually typical. Thus, if $S_2'$ is another s-set such that $S_2' \cap S_2 \in \cal S$ and $||\Psi(S_2 \cap S_2')||^2$ is not ``too small'' relative to $||\Psi(S_2)||^2$, we expect that $P(S_1 \cap S_2') \approx P(S_2 \cap S_2')=||\Psi(S_2 \cap S_2')||^2$.

This result can be proven rigorously. In fact, for all $P \in {\cal P}_\Psi$, we have the inequality:
\begin{equation} \label{ii}
||\Psi(S_2 \cap S_2')||^2 - ||\Psi(S_1)- \Psi(S_2)||^2 \leq P(S_1 \cap S_2') \leq ||\Psi(S_2 \cap S_2')||^2 + ||\Psi(S_1)- \Psi(S_2)||^2,
\end{equation}
for $||\Psi(S_1)||=||\Psi(S_2)||$ and $S_2 \cap S_2' \in \cal S$. Thus, if $S_1$, $S_2$ and $S'_2$ are defined as above, we have
\begin{equation}
\frac{||\Psi(S_1) - \Psi(S_2)||^2}{||\Psi(S_2 \cap S_2')||^2} \ll 1,
\end{equation}
and therefore 
\begin{equation}
\frac{\left | P(S_1 \cap S_2') - ||\Psi(S_1 \cap S_2')||^2 \right |}{||\Psi(S_2 \cap S_2')||^2} \ll 1.
\end{equation}

Let us prove inequality (\ref{ii}). We have
\begin{eqnarray*} 
& & P(S_1 \cap S_2')  \geq P(S_1 \cap S_2 \cap S_2') = P(S_1 \cap S_2) - P(S_1 \cap S_2 \cap {S_2'}^c) \geq \\
& & ||\Psi(S_2)||^2 - ||\Psi(S_1) - \Psi(S_2) ||^2 - P(S_2 \cap {S_2'}^c) = ||\Psi(S_1 \cap S_2')||^2 - ||\Psi(S_1) - \Psi(S_2) ||^2.
\end{eqnarray*}
Moreover 
\begin{eqnarray*}
& & P(S_1 \cap S_2') = P(S_1 \cap S_2' \cap S_2) + P(S_1 \cap S_2' \cap {S_2}^c) \leq P(S_2' \cap S_2) + P(S_1 \cap {S_2}^c) = \\
& & ||\Psi(S_2 \cap S_2')||^2 + P(S_1) - P(S_1 \cap S_2) \leq ||\Psi(S_1 \cap S_2')||^2 + ||\Psi(S_1) - \Psi(S_2) ||^2.
\end{eqnarray*}

\vspace{3mm}
Let us now refer to another property. In \cite{galvan} and in section \ref{int} we mentioned that the trajectories follow approximately the branches of the wave function. We can now give a precise mathematical formulation of this assertion.

Let us consider again the wave packets $\phi(t)$ and $\phi_\perp(t)$ defined above, and let us suppose that they do not overlap during the entire time interval $[t_1, t_2]$. The wave packet $\phi(t)$ in the time interval $[t_1, t_2]$ is what we view as a {\it branch} of the wave function. For $t \in [t_1, t_2]$ let $\Delta_t$ be the support of $\phi(t)$, with $||E(\Delta_{t})\Psi(t)||=||E(\Delta_{t_1})\Psi(t_1)||$ for $t \in [t_1, t_2]$, and let $S_t$ and $S_1$ denote the s-sets $(t, \Delta_t)$ and $(t_1, \Delta_{t_1})$ respectively.  According to the reasoning of section \ref{quantum}, we then have
\begin{equation}
\frac{||\Psi(S_1)- \Psi(S_{t})||^2}{||\Psi(S_1)||^2} \leq \epsilon \ll 1 \; \; \hbox{for all} \; \; t \in [t_1, t_2],
\end{equation}
and therefore, according to the quantum typicality rule:
\begin{equation}
\frac{P(S_1 \cap S_t)}{P(S_1)} \geq 1 - \epsilon \; \; \hbox{for all } \; \; t \in [t_1, t_2] \; \; \hbox{and} \; \; P \in {\cal P}_\Psi.
\end{equation}

Let $\{s_1, \ldots, s_n \}$ be any sequence of times in the time interval $[t_1, t_2]$. Moreover, for any $P \in {\cal P}_\Psi$, let $(S_1, {\cal F} \cap S_1, P( \cdot | S_1))$ be the probability space obtained from $(X^T, {\cal F}, P)$ by restricting $X^T$ to $S_1$. On this space let us introduce the random variable $Y:S_1 \rightarrow [0,1]$ defined by:
\begin{equation}
Y(\lambda):=\frac{1}{n}\sum_{i=1}^n {\bf 1}_{\Delta_{s_i}}[\lambda(s_i)].
\end{equation}
One can show that 
\begin{equation} \label{w11}
E_P(Y)\geq 1 - \epsilon \; \; \hbox{and} \; \;  P(Y \leq 1 - \delta) \leq \frac{\epsilon}{\delta} \; \; \; \; \hbox{for all} \; \; P \in {\cal P}_\Psi,
\end{equation}
where $E_P(Y)$ is the expectation value of $Y$ (the dependence on the probability measure has been explicitly shown) and $\delta$ is a suitable ``small'' positive number. Indeed we have
$$
E_P(Y)=\frac{1}{P(S_1)}\int_{S_1} Y \, dP=\frac{\sum_i  P(S_1 \cap S_{s_i})}{ n \, P(S_1)} \geq 1 - \epsilon.
$$
As to the second inequality (\ref{w11}), let $a$ be a given point of the interval $[0,1]$ and $0 \leq P_a \leq 1$ a given value (of probability). We have 
$$
\sup_{\{Y: P(Y \leq a )=P_a \} } \left \{E_P(Y) \right \} = a  P_a + (1 - P_a)=1 -P_a(1 - a).
$$
Indeed the supremum of the expectation value is obtained when the probability density $\rho(y)$ of $Y$ is shifted as much as possible on the right of the interval $[0,1]$ compatibly with the constraint  $P(Y \leq a)=P_a$, that is when it is of the form $\rho(y)=\delta(y -a) P_a + \delta(y - 1) (1- P_a)$. The second inequality (\ref{w11}) is obtained by posing the condition  $1 -P_a(1 - a) \geq 1 - \epsilon$ and by replacing $a$ with $1 - \delta$.

\vspace{3mm}
Conditions (\ref{w11}) hold true for any sequence of times. By choosing a ``dense'' sequence, the value of $Y(\lambda)$ can be assumed to correspond to the fraction of the time interval $[t_1, t_2]$ that the trajectory $\lambda$ spends inside $\Delta_t$. Thus, if for example $\epsilon = 10^{-6}$, from the second inequality (\ref{w11}) we obtain
\begin{equation}
P(Y \leq 1 - 10^{-3}) \leq 10^{-3}.
\end{equation}
In other words, the overwhelming majority of the trajectories of $S_1$ spend the greatest part of the time interval $[t_1, t_2]$ inside the support $\Delta_t$. This is the mathematical formulation of the assertion that the trajectories follow approximately the branches of the wave function.

\section{Discussion and conclusion} \label{discussion}

Let us summarize the formulation of quantum mechanics as a theory of imprecise probability.

According to this formulation, a closed system of quantum particles is represented as a quantum process $(X^T, {\cal F}, {\cal P}_\Psi)$, where $X$ is the configuration space of the particles, $T$ is a time interval, ${\cal F}$ is the $\sigma$-algebra generated by the s-sets (or equivalently, by the cylinder sets) and ${\cal P}_\Psi$ is the set of probability measures on $X^T$ defined by the formal quantum typicality rule:
\begin{equation}
P(S_1 \cap S_2) \geq ||\Psi(S_1)||^2 - ||\Psi(S_1) - \Psi(S_2)||^2 \; \; \hbox{for} \; \; ||\Psi(S_1)||=||\Psi(S_2)||,
\end{equation}
where $S_i=(t_i, \Delta_i)$ and $\Psi(S_i)=U^\dagger(t_i)E(\Delta_i)U(t_i)\Psi_0$.

The meaning of this interpretation is that the trajectory followed by the quantum particles corresponds to a trajectory chosen at random (according to any one of the probabilities of ${\cal P}_\Psi$) from the set $X^T$. The predictive power of this model is limited to those events $A$ for which $P(A)$ has approximately the same value for all $P \in {\cal P}_\Psi$. This is true, for example, for the s-sets, for which $P(S)=||\Psi(S)||^2$, and for the intersections of non equal-time s-sets satisfying the condition described in section \ref{two}. As we will now discuss, this appears to be sufficient to explain the results of the statistical experiments and the quasi-classical structure of the macroscopic evolution.

Since the only system which is really closed is the universe, the basic assumption is that the entire universe is represented as a quantum process. The fact that subsystems of the universe can also be represented as quantum processes would have to be deduced from the basic assumption by means of reasoning analogous to that adopted in \cite{durr1}. Here, we do not make this reasoning explicit. Since there is just one universe, just one trajectory is chosen from $X^T$, and its properties derive from reasoning based on typicality rather than on probability: suppose that $A \in {\cal F}$ is the set of the trajectories satisfying a suitable property, and that $P(A) \approx 1$ for all $P \in {\cal P}_\Psi$. This explains why a single trajectory chosen at random from $X^T$ satisfies the property \cite{durr1, goldstein}. For example, we have seen in section \ref{two} that the overwhelming majority of the trajectories of $X^T$ follows approximately the branches of the universal wave function.

This formulation of quantum mechanics is a {\it trajectory based formulation}, analogous for example to Bohmian mechanics or to Nelson's stochastic mechanics. In these formulations the particles follow definite trajectories, their positions are the only observable quantities and neither the measurement process nor the observers enter into the theory on a fundamental level. The standard quantum measurement theory can be derived in this context on the basis of the fact that any measurement performed in a real laboratory ultimately comes down to a measurement of the position of a pointer \cite{operators}.

A quantum process arguably explains both the results of the statistical experiments and the quasi-classical  structure of macroscopic evolution. The former explanation derives from the fact that all the probabilities of ${\cal P}_\Psi$ satisfy the Born rule.

The latter explanation derives from the assumption that the universal wave function has a branching structure, i.e. that it can be split at any time into permanently non-overlapping wave packets, and that the supports of the wave packets have a limited extension, i.e. an extension compatible with a well defined macroscopic configuration. When, during the time evolution, the support of a wave packet extends over a region no longer compatible with a well defined macroscopic configuration, it is assumed that it can be further split into smaller wave packets with the required extension. Moreover, the macroscopic configurations corresponding to the supports of the branches (or at least, to the overwhelming majority of them) are assumed to evolve quasi-classically. The branching structure of the universal wave function is accepted by various authors \cite{bohm, struyve}. The quasi-classical evolution of the branches would have to be derived from the Ehrenfest theorem and/or from reasoning analogous to Mott's analysis of the bubble chamber experiment \cite{mott}. Of course, all these assumptions require a more rigorous investigation. Note that, in this formulation of quantum mechanics, the nature of the branching process is well defined, namely a branch is a {\it spatially} non-overlapping wave packet. This is not the case, for example, in the Many Worlds interpretation, in which the preferred-basis problem appears to be still open. Given the branching structure for the universal wave function described above, the quasi-classical structure of the trajectory of our universe then derives from the fact that the overwhelming majority of the trajectories follow approximately the branches of the wave function, as shown in section \ref{two}.

A last remark regarding the role played by the set $X^T$ in a quantum process: usually, in quantum mechanics position variables are considered ``hidden variables''. In this case, however, $X^T$ is the sample space of an (imprecise) probability space, and therefore such a definition appears to be inappropriate. The set $X^T$ has no empirical content, i.e. no empirical prediction can be derived from it. On the contrary, all the predictions can be derived from the set ${\cal P}_\Psi$ alone, that is from the wave function. However $X^T$ cannot be removed from a quantum process, for the same reason for which the sample space cannot be removed from a probability space. In other words, without the set $X^T$ we can calculate everything, but we have serious coherence problems. This situation appears to reflect the current situation of quantum mechanics.

\end{document}